\documentclass[twocolumn,preprintnumbers,aps]{revtex4}
\usepackage{graphicx}

\newcommand{\bi}{\bibitem}
\newcommand{\be}{\begin{eqnarray}}
\newcommand{\ee}{\end{eqnarray}}
\newcommand{\rar}{\rightarrow}

\begin{document}

\title{Dangerous implications of a minimum length in quantum gravity}

\author{Cosimo Bambi$^{\rm 1, 2}$}
\author{Katherine Freese$^{\rm 2}$}

\affiliation{$^{\rm 1}$Department of Physics and Astronomy, 
Wayne State University, Detroit, MI 48201, USA\\
$^{\rm 2}$Michigan Center for Theoretical Physics, University of Michigan,
Ann Arbor, MI 48109, USA}

\date{\today}

\preprint{WSU-HEP-0802}

\begin{abstract}
The existence of a minimum length and a generalization of the
Heisenberg uncertainty principle seem to be two fundamental
ingredients required in any consistent theory of quantum gravity. 
In this letter we show that they would predict dangerous 
processes which are phenomenologically unacceptable. For 
example, long--lived virtual super--Planck mass black holes may 
lead to rapid proton decay. Possible solutions of this puzzle 
are briefly discussed.
\end{abstract}

\maketitle

{\sc Introduction ---} 
The search for a theory of quantum gravity is a sort of
Holy Grail in physics; yet, despite a great deal of effort, 
even the very meaning of quantum spacetime is still not 
clear. The difficulties are essentially related to the quite 
peculiar nature of the gravitational force. At present 
we can at most rely on a small set of model--independent 
features which are expected to hold in any consistent 
framework of quantum gravity. In particular, from quite 
general grounds, we can deduce the existence of a minimum 
length and a generalization of the Heisenberg uncertainty 
principle of quantum mechanics~\cite{gup} (for a review, 
see ref.~\cite{garay}). The argument is basically the 
following. In quantum mechanics, the accuracy on the 
position of a particle is limited by the uncertainty on 
its momentum by the well known rule $\Delta x \gtrsim 1 / \Delta p$. 
On the other hand, in general relativity a certain amount 
of energy cannot be localized in a region smaller than the
one defined by its gravitational radius, so 
$\Delta x \gtrsim L_{Pl}^2 \Delta p$, where 
$L_{Pl} = G_N^{1/2} \sim 10^{-33}$~cm is the Planck length. 
Now, combining the two results, we conclude that there 
must exist a minimum observable length
\be
\Delta x \gtrsim \max 
\left( \frac{1}{\Delta p} ; L_{Pl}^2 \Delta p \right) 
\gtrsim L_{Pl} \, .
\ee
Such a result can be summarized in the so called Generalized 
Uncertainty Principle, which can be deduced from more 
sophisticated gedanken experiments in different 
frameworks~\cite{gup} and reads
\be\label{eq-gup}
\Delta x \gtrsim \frac{1}{\Delta p} 
+ \alpha L_{Pl}^2 \Delta p \, ,
\ee
where $\alpha$ is some positive and dimensionless 
model--dependent coefficient, usually assumed to be of order one.
As one can easily suspect at this point, it is also
possible to derive a generalized time--energy uncertainty
relation~\cite{garay, time}
\be\label{eq-gup-time}
\Delta t \gtrsim \frac{1}{\Delta E} 
+ \beta L_{Pl}^2 \Delta E \, ,
\ee
where $\beta$ is a new positive model--dependent parameter.

{\sc The paradox ---}
In quantum mechanics, the nature of the relation between time 
and energy $\Delta t \Delta E \gtrsim 1$ is quite different 
from the one between position and momentum $\Delta x \Delta p \gtrsim 1$: 
$t$ is a parameter and not an operator, the corresponding 
formula does not arise from ``first principles'' (i.e. from 
the commutation relation) and in the end it is not always true, 
see e.g. the discussion in ref.~\cite{nikolic}. Nevertheless, 
in most physical processes it works and is capable of providing 
a simple interpretation of several phenomena. For example, it 
provides a mechanism to evaluate the lifetime of an excited 
state from its energy width in atomic and nuclear physics. In 
quantum field theory, it allows for an interpretation according 
to which a virtual particle of energy $E$ can be produced out 
of vacuum for a time $1/E$. Even if we still do not have any 
reliable theory of quantum gravity, we can expect that such 
general considerations are common to any quantum theory and 
are not ruled out by including the effects of gravitational 
interaction (or at least this is the only thing we can do at 
present!). Hence, if we take eq.~(\ref{eq-gup-time}) seriously, 
we are induced to assert that in quantum gravity any excited 
state of energy $E$ can be produced out of vacuum for a time
\be
\tau \sim \max\left(\frac{1}{E} ; L_{Pl}^2 E \right) \, .
\ee
The result is that super--Planckian excitations can exist for a 
very long time.  This fact presents a paradox, because it implies
that the contribution from heavier and heavier states of the
theory should be more and more important, even in low energy
physical processes.  Such a situation destroys the success of 
our current framework of particle physics and predicts even 
more dangerous events, which are clearly inconsistent with the
universe in which we live.

One might worry about the contribution of long--lived virtual
(1) classical macroscopic objects, (2) sub--Planck mass particles
with super--Planckian energies, or (3) super--Planck mass particles
to dangerous physical processes; we will show that it is the 
third category that is really problematic.

In principle, eq.~(\ref{eq-gup-time}) could permit  
classical macroscopic virtual objects to live for very 
long times, because their mass is much larger than the 
Planck mass $M_{Pl} \sim 10^{-5}$~g. However, such a 
possibility is not dangerous, because classical objects 
are extremely tidy states, so they should be suppressed 
by a huge entropy factor. This suppression is familiar
in a different context: in particle physics accelerators, 
it is extremely unlikely that large classical objects 
such as magnetic monopoles or sphalerons could be produced,
though a rigorous calculation has not been 
performed~\footnote{For example, monopole--antimonopole 
annihilation into all possible particles should be quite 
efficient, with a cross section $\sigma \sim (m_M)^{-2}$ and a 
reaction probability $\Gamma \sim m_M$, where $m_M$ is the 
monopole mass and the only energy scale in the problem. One 
might expect the inverse process, the assembly of multiple 
particles into a monopole--antimonopole pair to proceed with
the same probability.  However, the number of particles
produced by monopole--antimonopole annihilation is expected 
to be of order $1/\alpha \sim 100$ and in a very coherent 
state; such a coherent state is not likely ever to be created
in a particle collision, so that the production of monopoles
in accelerators is very unlikely ever to take place.}.
For this reason we can neglect the contribution of virtual
classical objects: the probability to produce them is not
determined mainly by the uncertainty principle, but by a huge
entropy factor which makes their creation hard even when the
process is classically allowed. For more details on the point, 
see e.g.~\cite{dolgov} and references therein. One could also
argue that the crucial ingredient of the generalization of the
Heisenberg uncertainty principle is the energy density, rather
than the total mass~\cite{sabine1}, and the one of macroscopic 
objects is typically much smaller than the ``critical value''
which is necessary for the formation of a horizon.

Secondly, one might worry about the contribution of virtual 
sub--Planck mass particles with energy beyond the Planck scale 
to dangerous physical processes. Here quantum field theory 
surely breaks down, computations become unreliable, and it 
would be difficult to predict the implications. However, as is 
usually done, we could solve the problem by asserting that the 
Planck scale represents a cut--off for the energy of an elementary 
particle. For example, in some frameworks the Planck energy and/or the 
Planck momentum cannot be reached by elementary objects~\cite{dsr}.
If this does not occur, the modified measure in momentum space
can do the job, acting as a soft cut--off: basically, 
eqs.~(\ref{eq-gup}) and (\ref{eq-gup-time}) can be seen as a 
redefinition of the Planck constant $h$, which sets the infinitesimal
volume of the phase space. This affects loop integration with a
correction of the form~\footnote{The exact expression is however 
model--dependent: it is determined by the canonical commutation 
relation $[x,p]$ which are not unique, i.e. there are infinite
choices leading to eq.~(\ref{eq-gup}).}
\be
\int \frac{d^4 p}{(2 \pi)^4} \rar \int 
\frac{d^4p}{(2 \pi)^4 \, \left(1 + \alpha' L_{Pl}^2 p^2\right)^4}
\ee
and suppresses high energy contributions.

The true paradox is instead related to the contribution of 
super--Planck mass virtual black holes: since in any 
quantum theory one expects to find the fundamental excitations 
of the theory, virtual black holes are supposed to be present 
in the mass spectrum of the (yet unknown) theory of quantum 
gravity. The standard argument for them is the spacetime foam
picture~\cite{foam1, foam2}. Using a semiclassical approach,
one considers the path integral for $N$ non--interacting 
black holes and finds
\be
Z \sim \int_0^\infty dM \sum_{N=0}^{\infty}
\frac{1}{N!} \left(\frac{V}{L_{Pl}^3}\right)^N
\exp\left(-4 \pi L_{Pl}^2 N M^2\right) \, ,
\ee
where $V$ is a normalization volume. Let us note that the 
integration is over all the black hole masses, that is 
from zero to infinity. Nevertheless, the contribution of 
heavier and heavier black holes is more and more suppressed 
and since $L_{Pl}$ acts as natural cut--off, one eventually 
finds that the spacetime is essentially filled with Planck 
mass black holes which live for about one Planck time.

Assuming the validity of the Generalized Uncertainty Principle, 
the picture changes dramatically. Indeed, in any quantum 
theory at zero temperature, the contribution of virtual states 
is suppressed only by their limited existence, which is controlled
by the time--energy uncertainty relation and goes to zero in 
the classical limit. Since the Generalized Uncertainty Principle 
says that, beyond the Planck energy, the violation of a larger 
and larger amount of energy is more and more unsuppressed, the 
contribution of super--Planck mass virtual black holes is 
presumebly more important than the contribution of Planck size ones.
Heavy (i.e. with a mass $M_{BH} \gg M_{Pl}$) Schwarzschild black 
holes, and more generally heavy non--extremal black holes, might 
still be considered as semiclassical objects, exponentially 
suppressed by a huge factor (the point is however controversial, 
see e.g. ref.~\cite{bh-p}). 
It is instead common belief that extremal black holes can be 
thought of as ordinary particles~\cite{ebh-p}. For example, 
extremal Reissner--Nordstr\"om black holes have vanishing Hawking 
temperature, are expected to be stable, and can contribute here. 
The modified measure in momentum space cannot solve the problem, 
just because it is not the ingredient which determines the density 
of black hole states and it is not clear how a minimum length 
scale can act as cut--off in this case. So, if we take literally 
eq.~(\ref{eq-gup-time}), it seems very hard to prevent pair 
production of virtual extremal Reissner--Nordstr\"om black holes. 
Then, since black holes can violate global quantum numbers such 
as baryon and lepton number~\cite{dejan}, ordinary matter should 
be unstable and quickly decay.

{\sc Proton decay ---}
An unstable particle of mass $m$ usually decays into lighter
particles with a lifetime $\sim (g^2 m)^{-1}$, where $g$ is the
coupling constant of the interaction responsible for the decay. 
This is for example the case of the $t$--quark, which decays 
into real $b$--quark and $W$ boson with a lifetime $(\alpha m_t)^{-1}$.
On the other hand, if the process is kinematically forbidden,
it proceeds through the emission of virtual particles and is
therefore suppressed, making the particle lifetime increase.
More precisely, if the process requires an intermediate heavy
particle of mass $M$, it needs to violate energy conservation
for a time $t \sim 1/M$ in a volume $V \sim 1/M^n$, where $n$
is the number of space dimensions. Using an effective quantum
field theory, the process is described by a non--renormalizable
operator suppressed by some power of $1/M$, depending on the 
number of dimensions of the spacetime and on the kind of particles 
involved. This is quite a general rule which is independent of 
the interaction involved.

When gravity is taken into account, Planck mass virtual black 
holes are expected to induce proton decay  (see 
e.g.~\cite{pdecay1, pdecay2} and references therein). In an
effective field theory in 3+1 dimensions, the process is 
described in the standard picture by a dimension six operator, 
since two quarks of the proton are converted into a quark 
(or an antiquark) and a charged lepton:
\be\label{eq-d6o}
\mathcal{O}_6 \sim \frac{\psi\psi\psi\psi}{M_{Pl}^2} \, .
\ee
The estimated proton lifetime in the standard framework 
(that is with the standard uncertainty principle) is
\be\label{eq-pdecay}
\tau_p \sim \frac{M_{Pl}^4}{m_p^5} 
\sim 10^{45} \; {\rm yr}
\ee
and is consistent with present experimental limits, where 
several channels are bounded by $\tau_p \gtrsim 10^{33}$~yr.

Of course, the prediction in eq.~(\ref{eq-pdecay}) must be revised in
the case of the generalized uncertainty relations.  Now, beyond the
Planck scale, position and time uncertainties increase as the energy
increases. The old suppression factors, proportional to some power of
the inverse of the energy scale of the reaction (in the case of Planck
mass black holes proportional to some power of $1/M_{Pl}$), should be
replaced by the same power of the black hole gravitational radius
$M_{BH}/M_{Pl}^2$, where $M_{BH}$ is the black hole mass (which is now
larger than $M_{Pl}$). We can therefore guess that the proton lifetime
should be 
\be
\label{eq-pdecay-new} 
\tau_p \sim \frac{M_{Pl}^8}{m_p^5  \, M_{BH}^4} \, .  \ee 
This means that excited states heavier than
roughly $10^3 \, M_{Pl}$ would induce proton decay which is far too
rapid to be consistent with experiments.

Of course, the validity of an effective field theory to 
describe objects heavier than the Planck mass may be 
questionable. However we note that the semiclassical 
approach suggests that virtual black holes of every mass
are produced in the spacetime foam and can thus give a
non--zero contribution to the amplitude of physical processes. 
On the other hand, from the Generalized Uncertainty Principle 
we can argue that the contribution of super--Planck mass virtual
black holes have to be more important than the one of Planck 
mass black holes, because of the longer lifetime. Even if we do
not have a rigorous framework, semiclassical results are 
usually believed to be at least capable of providing the 
basic features. Surely the description breaks 
down for $M_{BH} \rar +\infty$, but we will show in the next
section that the rapid proton decay is mediated by virtual black 
holes for which the effective theory is likely still reliable. 
Lastly, we stress that the apparent paradox is a peculiar  
implication of the unsuppressed contributions of very heavy 
states. The problems of infinite pair production which are present
in attempts to incorporate black hole remnants in an effective 
theory~\cite{remnants1} are very different: in that case, the 
origin of the problems is due to the concept of remnant, because
remnants can be a viable solution to the black hole information 
paradox only if there is an infinite amount of states below 
some finite mass (see however ref.~\cite{remnants2} for a different 
point of view). Here we do not need the existence of remnants and,
without the Generalized Uncertainty Principle, our description 
of virtual black holes would be phenomenologically acceptable.

{\sc Possible solutions ---} 
Even if eq.~(\ref{eq-gup-time}) is deduced from heuristic
considerations, the arguments supporting the picture sound
quite robust and the new position--momentum and time--energy 
uncertainty relations look at least reasonable. 
In this case it is not easy to suppress the contribution of 
virtual black holes: the simple replacement of the Heisenberg
uncertainty principle with the Generalized Uncertainty Principle
leads to a paradox and we should figure out what is wrong.  
Before proceeding, we would like to
stress that these black holes are always ``virtual'', that is cannot
be seen directly, since we cannot produce the super--Planckian energies
required to create them. Nevertheless, they should be observed
indirectly, through their contribution in low energy physical events.

Broadly speaking, an elegant mechanism which is capable of forbidding
(if exact) or reducing considerably (if approximate) dangerous
processes otherwise inevitable is the introduction of a new
symmetry. However, this is not so easy here: a global symmetry is
expected to be violated by black holes, while an unbroken gauge
symmetry demands a new massless gauge boson, which is questionable at
best or more likely phenomenologically forbidden.  Broken gauge
symmetries also cannot work, because a charge related to the gauge
boson of mass $m$ disappears without any trace inside the black hole,
with a characteristic time $1/m$~\cite{bh-hairs}; i.e., the quantum
number associated with the symmetry is violated.  At present, the only
known possibility is to introduce a gauge symmetry which is broken to
a discrete remnant~\cite{krauss}: in this case, even if the gauge
boson acquires a mass, Gauss's law implies that when a charge related
to the symmetry falls into the black hole, a surface integral which is
an intrinsic feature of the asymptotic black hole state still survives
and the quantum number is unchanged.  An explicit realization can be
found in ref.~\cite{faraggi}, while for more general considerations
see ref.~\cite{pawl}.

Another possibility could be to appeal to the peculiar (and not well
understood) nature of non--extremal and extremal black holes: the
former might be exponentially suppressed because they are
semiclassical objects, the latter might be unable to violate global
quantum numbers. For instance, although virtual long-lived extremal
Reissner--Nordstr\"om black holes can come into existence, one might
argue (based on the fact that they have zero temperature) that these
black holes do not violate baryon number~\cite{dejan}. However, such
an explanation cannot work for other extremal states: examples include
the extremal black holes of the charged dilaton family, which can have
non--zero temperature even if zero entropy. On the other hand, these
extremal states may not exist at all, because we do not know if
dilaton--like fields really exist in nature.

However, all these proposed solutions can at best avoid too rapid
proton decay and other dangerous baryon or lepton violating processes.
There is still the danger that heavy extremal black holes can 
participate in ordinary reactions in the Standard Model
of particle physics that are thus unsuppressed. Then the branching
ratios of ordinary processes would be altered, leading to disagreement
with experiment.
All the observed branching ratios are essentially in agreement with 
theoretical predictions, while black holes should change them, because 
the effective field theory we have outlined in this paper breaks 
down for very heavy black holes. To be more quantitative, we can
consider a general process which is allowed in the Standard Model, 
for instance $e + \nu_\mu \rar \mu + \nu_e$. Such a process should 
be possible with an intermediate virtual black hole as well and in 
that case it would be described by a dimension 6 operator like the 
one in eq.~(\ref{eq-d6o}). In the standard theory, $M$ is the Planck
mass and the differential cross section is
\be
\frac{d\sigma}{d\Omega} \sim \frac{E^2}{M_{Pl}^4}  
\sim | f(\theta) |^2 \, ,
\ee
where $f(\theta)$ is the scattering amplitude. On the other hand,
in the case of the Generalized Uncertainty Principle 
$M = M_{Pl}^2 / M_{BH}$ and the differential cross section becomes 
\be
\frac{d\sigma}{d\Omega} \sim \frac{E^2 M_{BH}^4}{M_{Pl}^8}
\sim |f(\theta)|^2 \, .
\ee
The scattering amplitude can be expanded in partial waves
\be
f(\theta) = \frac{1}{E} \sum_{l=0}^{\infty} 
\left( l + \frac{1}{2} \right) M_l P_l(\cos\theta)
\ee
and unitarity demands $|M_l| \le 1$ for any $l$. For $l = 0$,
we have
\be
f(\theta) = \frac{M_0}{2 E}
\ee
and the unitarity bound reads
\be
\frac{4 E^4 M_{BH}^4}{M_{Pl}^8} \sim |M_0|^2 \le 1 \, .
\ee
This means that for a process with a characteristic energy scale 
$E \sim 1$~GeV, unitarity is lost for black holes heavier than 
$M_{max} \sim M_{Pl}^2/E \sim 10^{15}$~g.

So, if we want to prevent any possible participation of virtual
black holes in low energy physics, we need a more convincing 
explanation, since mechanisms like a new symmetry are unable to 
do the job. Generally speaking, one could expect that a minimum length
scale acts as a regulator of the theory and somehow is capable of
suppressing all the dangerous contributions above the Planck
scale~\cite{sabine2}.  The realization of such a mechanism is yet to be
investigated, and, unfortunately, it is unlikely that we are able to
give a final answer; we would need a theory capable of providing clear
predictions at the Planck scale, while at present we can at best make
vague conjectures.   
A quite peculiar mechanism that drastically reduces any B and L
violating process in the case of theories with a low gravity scale at
the level to be consistent with experimental bounds has been proposed
in ref.~\cite{pdecay2} and can work even here. There the
conjecture is that black holes lighter than the (effective) Planck
mass must have zero charge and zero angular momentum in quantum
gravity as well as classically so that proton decay is drastically
suppressed.  It is conjectured that black holes
cannot be produced out of vacuum in violation of energy conservation,
even for very short times (e.g. in 1 GeV processes only 1 GeV black
holes can be produced, not very heavy ones).  If we then include the
idea of a minimum length, which implies a minimum black hole mass at
the level of the Planck mass, black holes cannot play any role in
particle physics at energies below the Planck scale.
Extra dimensions might also be a solution by
allowing for (otherwise unnaturally) small coupling
constants~\cite{arkani}.

{\sc Conclusion ---} 
Although a reliable theory of quantum gravity is still lacking, 
there are some basic ingredients (such as a minimum observable 
length, the Generalized Uncertainty Principle and the idea of
virtual black holes) which are commonly believed to be present
in the theory of quantum gravity. It is not easy to reject the 
conjectures that eq.~(\ref{eq-gup-time}) holds and that 
super--Planck mass virtual black holes predict naturally
dangerous processes, clearly inconsistent with the observed 
universe. In particular, rapid baryon number violating processes 
may lead to predictions of proton decay lifetimes that are ruled 
out by experiment. Even if B and L violating processes are 
somehow forbidden, there are other problems caused by these
super--Planck mass virtual black holes: heavier and heavier black 
holes are more and more unsuppressed, and we still have to 
explain how they do not alter branching ratios predicted by the
Standard Model of particle physics. We would like to stress that 
such dangerous processes are expected in the constructed effective 
theory for virtual black holes that do not violate unitarity and 
hence the prediction is presumebly reliable. The problem is even
more evident if we recall the peculiar predictions that can be
easily deduced from the standard uncertainty principle (stability
of atoms, lifetime of unstable states, stability of degenerate
stars, etc.): if the idea of the Generalized Uncertainty Principle
and of the spacetime foam picture are basically correct, why cannot
we put them together and believe in their predictions?

In this letter we have formulated a problem which demands an 
answer. In particular, if we do not want to reject the Generalized 
Uncertainty Principle, we have to figure out the reason for 
which dangerous predictions cannot occur. One possibility is 
to solve the puzzle with common ingredients in particle physics, 
say the introduction of a symmetry or something similar. The 
most radical case is instead that there is something wrong in 
what we believe to find in the theory of quantum gravity.

We believe that the study of the present problem may shed new light on
the search for a theory of quantum gravity and for a unified
description of gravitational and non--gravitational interactions. In
addition to this, the emerging picture may be also interesting from
the experimental point of view, allowing for a proton lifetime not too
far from present experimental bounds.  The standard description of
black holes is usually expected to break down below some minimum mass
$M_{min}$, which is not too much larger than $M_{Pl}$. In this case,
we can imagine the situation in which there is a suppression mechanism
that works for dangerous heavy black holes, yet may not hold for
purely quantum mechanically objects just one or two orders of
magnitude heavier than the Planck mass. For example, if the
contribution from black holes with a mass $M_{BH} \sim 10^2 - 10^3 \;
M_{Pl}$ is not too suppressed, the proton lifetime
could be on the verge of present experimental limits.

\begin{acknowledgments}
We wish to thank Alexander Dolgov, David Garfinkle, Sabine
Hossenfelder, Tsvi Piran and Dejan Stojkovic for useful comments 
and suggestions.
C.B. is supported in part by NSF under grant PHY-0547794 
and by DOE under contract DE-FG02-96ER41005 and thanks
the MCTP for hospitality during his visit.  K.F. acknowledges
support from the DOE and the MCTP via the University of Michigan.
\end{acknowledgments}


\end{document}